# The Design and Architecture of the Microsoft Cluster Service
*-- A Practical Approach to High-Availability and Scalability*

Werner Vogels[1],

Dan Dumitriu[1],

Ken Birman[1],

Rod Gamache,

Mike Massa,

Rob Short,

John Vert

Joe Barrera,

Jim Gray

May 1998

cs.OS/9809006   2 Sep 1998

---

[1] *Dept. of Computer Science, Cornell University*

# The Design and Architecture of the Microsoft Cluster Service
*-- A Practical Approach to High-Availability and Scalability* [†]

Werner Vogels,
Dan Dumitriu, Ken Birman
*Dept. of Computer Science*
*Cornell University*

Rod Gamache, Mike Massa,
Rob Short, John Vert
*Microsoft Cluster group*
*Microsoft Corporation*

Joe Barrera,
Jim Gray
*Scalable Server group*
*Microsoft Research*

## Abstract

*Microsoft Cluster Service (MSCS) extends the Windows NT operating system to support high-availability services. The goal is to offer an execution environment where off-the-shelf server applications can continue to operate, even in the presence of node failures. Later versions of MSCS will provide scalability via a node and application management system which allows applications to scale to hundreds of nodes. In this paper we provide a detailed description of the MSCS architecture and the design decisions that have driven the implementation of the service. The paper also describes how some major applications use the MSCS features, and describes features added to make it easier to implement and manage fault-tolerant applications on MSCS.*

## 1 Introduction

A cluster is a collection of computer nodes that work in concert to provide a much more powerful system. To be effective, the cluster must be as easy to program and manage as a single large computer. Clusters have the advantage that they can grow much larger than the largest single node, they can tolerate node failures and continue to offer service, and they can be built from inexpensive components.

Cluster computing is poised to surge in importance with the emergence of software that supports scalable clusters using commodity components. Traditionally, cluster architectures relied on special-purpose hardware. Software clusters eliminate the need for proprietary hardware. A software cluster can scale to many nodes at a single site, and can scale *geographically*, creating a single "server" that spans multiple locations. Software clusters can also offer improved management and ease-of-use. These benefits are well matched to a web-oriented computing model. Software clusters, integrated with tools for cluster application development will create new applicatons for both scalable and fault-tolerant systems.

For clusters to realize this promise, cluster technology must improve. Users find it difficult to configure clusters with the desired management and security properties. It is difficult to configure applications to be automatically launched in an appropriate order, to deal with wide-area integration issues, and otherwise to match the cluster to application needs. Lacking solutions to these problems, clusters will remain awkward and time-consuming tools, limiting the growth of *cluster-aware* applications.

Microsoft Cluster Service (MSCS) takes a phased approach to solving these problems. The first phase addresses high-availability file servers, databases, web servers, and electronic mail servers. For many businesses, these servers have become essential to daily operation. MSCS extends Windows NT™ with mechanisms to improve application availability. MSCS detects and restarts failed components, reducing the mean-time-to-repair (MTTR) . MSCS also migrates components to other nodes if one node of the cluster fails. Migration improves availability by more than an order of magnitude.

In this first phase, MSCS offers only minimal support for application scalability to two nodes. Later MSCS releases may support larger and geographically distributed clusters. They will also improve support for self-management of distributed applications, and for the development of cluster-aware (parallel) applications.

Microsoft Cluster Service is not the first technology to support failover, migration, and automated restart of failed components. Important prior work includes the application fail-over support available on many commercial cluster platforms, notably the DEC, HP, IBM, NCR, Tandem and Stratus failover products. See [5] for a more complete review of these and other cluster solutions. MSCS goes beyond prior work by providing a signifi-

---

[†] This paper is the result of a collaboration between the Reliable Distributed Computing group at the Computer Science Department of Cornell University, the Cluster Group at Microsoft Corporation and the Scalable Server Group at Microsoft Research. Cornell's Reliable Distributed Computing group is analyzing the distributed system components of MSCS and designing new algorithms for scalable operation and cluster-aware application programming. An extended version of this analysis paper is also available [6].

The research by the Reliable Distributed Computing Group is supported by DARPA/ONR under contract N0014-96-1-10014 and by Intel Corporation and Microsoft Corporation. Inquiries with respect to this paper or the full report, can be send to Werner Vogels, email: vogels@cs.cornell.edu.

cantly simpler use interface and greater sophistication in the way that applications are modeled. Moreover, MSCS has a tighter integration with the operating system than do most other cluster solutions.

The connections between cluster-style computing and prior work on reliable group management and communication (atomic multicast) are of interest. Tracking the active set of nodes in a cluster corresponds to the group membership problem [1]. Avoiding the "split brain syndrome" (whereby a cluster splits into two disjoint parts that both claim to own some critical resource) is analogous to the primary component network partitioning problem. Linking clusters into a geographically distributed wide-area system is similar to the wide-area process-group problem [1]. Maintaining a checkpoint and log for use during restart is an instance of the more general transaction processing techniques of logging and commit/abort to perform atomic state transformations on all the replicas [3]. A cluster can thus be viewed as a way to package powerful fault-tolerance primitives in a way that is natural and convenient for a very large class of users for whom application availability is a key objective.

The review of the Microsoft Cluster Service that follows focuses on the fundamental abstractions of the first phase. It describes the software architecture of the cluster service and describes the structure of some major commercial applications that use MSCS.

## 2 Cluster design goals

A cluster managed by the Microsoft Cluster Service is a set of loosely coupled, independent computer nodes, which presents a single system image to its clients. MSCS adopts a *shared nothing* cluster model, where each node within the cluster owns a subset of the cluster resources. Only one node may own a particular resource at a time, although, on a failure, another node may take ownership of the resource. Client requests are automatically routed to the node that owns the resource.

The first phase of MSCS, released in late 1997 had the following general goals:

- **Commodity**. The cluster runs on a collection of off-the-shelf computer nodes interconnected by a generic network. The operating system is a standard commercial version of Windows NT server, the network communication is through the standard Internet protocols.
- **Scalability**. Adding applications, nodes, peripherals, and network interconnects is possible without interrupting the availability of the services at the cluster.
- **Transparency**. The cluster, which is built out of a group of loosely coupled, independent computer nodes, presents itself as a single system to clients outside the cluster. Client applications interact with the cluster as if it were a single high-performance, highly reliable server. The clients as such, are not affected by interaction with the cluster and do not need modification. System management tools access and manage the services at the cluster as if it is one single server. Service and system execution information is available in single image, cluster wide logs.
- **Reliability**. The Cluster Service is able to detect failures of the hardware and software resources it manages. In case of failure the Cluster Service can restart failed applications on other nodes in the cluster. The restart policy is part of the cluster configuration. It can specify the availability requirements for that application. A failure can also cause ownership of other resources (shared disks, network names, etc.) to migrate to other nodes in the system. Hardware and software can be upgraded in a phased manner without interrupting the availability of the services in the cluster.

Several issues were explicitly **not** part of the first phase of the design: MSCS proves no development support for fault-tolerant applications (process pair, primary-backup, active replication), no facilities for the migration of running applications, and no support for the recovery of the shared state between client and server. However, all of these are viewed as options for futures design phases.

## 3 Cluster Abstractions

MSCS is designed around the abstractions of nodes, resources, resources dependencies, and resource groups. This section describes each of these central abstractions and the relations between the abstractions. The next section on Cluster Operation will provide the context in which the abstractions are used.

### 3.1 Node

A node is a self-contained Windows NT™ system that can run an instance of the Cluster Service. Groups of nodes implement a cluster. Nodes in a cluster communicate via messages over network interconnects. They use communication timeouts to detect node failures. There are two types of nodes in the cluster: (1) *defined* nodes are all possible nodes that can be cluster members, and (2) *active* nodes are the current cluster members. A node is in one of three states: *Offline*, *Online*, or *Paused* (see sections 4.1 and 5.1 for details on this).

### 3.2 Resource

A resource represents certain functionality offered at a node. It may be physical, for example a printer, or logical for example an IP address. Resources are the basic management units. Resources may, under control of the Cluster Service, migrate to another node.

MSCS implements several resource types: physical hardware such as shared SCSI disks and logical items such as disk volumes, IP addresses, NetBios names and SMB server shares. Applications extend this set by implementing logical resources such as web server roots, transaction mangers, Lotus or Exchange mail databases, SQL databases, or SAP applications.

Resources can fail. The Cluster Service uses resource monitors (section 4.2) to track the status of the resources. The cluster service restarts resources when they fail or when one of the resources they depend on fails.

A resource has an associated *type*, which describes the resource, its generic attributes, and the resource's behavior as observed by the Cluster Service. One of these attributes is a resource control library that is used by the resource monitors to implement the specific monitoring for the type of resource.

### 3.3 Quorum Resource

The *quorum resource* provides an arbitration mechanism to control membership. The quorum resource also implements persistent storage where the Cluster Service can store the Cluster Configuration Database and change log. The Quorum Resource must be available when the cluster is formed, and whenever the Cluster Configuration Database is changed. It is desirable that the Quorum resource be highly available and that it not depend on the availability of a single node. At present, MSCS employs a partition on a shared fault-tolerant SCSI disk to implement the Quorum Resource, although other technologies may be employed for this purpose in the future.

### 3.4 Resource Dependencies

Resources often depend on the availability of other resources. An instance of a SQL server depends on the presence of a certain SQL database that in turn depends on the availability of the disks that store the database. These dependencies are declared and recorded in a *dependency tree*. The dependency tree describes the sequence in which the resources should be brought online. It also describes which resources need to migrate together. If a resource is restarted, all resources that depend on it are also restarted. Dependencies cannot cross resource group boundaries.

### 3.5 Resource Groups

A Resource Group is the unit of migration (failover). Although a resource dependency tree describes the resources which **must** failover together, there may be additional considerations for grouping resources into migration units. The cluster administrator can assign a collection of independent resource dependency trees to a single resource group. When the group needs to migrate to another node in the cluster, all the resources in the group will move to the new location. Failover policies are set on a group basis, including the list of preferred owner nodes, the failback window, etc.

### 3.6 Cluster Database

All configuration data necessary to start the cluster is kept in the Cluster Configuration Database. The database, which is replicated at each node in the cluster, is accessed through the standard Windows NT configuration database, called the *registry*. The initial node forming the cluster initializes the database from the Quorum Resource, which stores the master copy of the database change logs. The Cluster Service, during the Cluster Form or Join operations, ensures that the replica of the configuration database is correct at each active node. When a node joins the cluster, it contacts an active member to determine the current version of the database and to synchronize its local replica of the configuration database. Updates to the database during the regular operation are applied to the Master copy and to all the replicas using an atomic update protocol similar to Carr's Global Update Protocol [2].

## 4 Cluster Operation

There are four areas of particular interest in an MSCS cluster: (1) cluster membership activities, (2) resource management and resource failure handling, (3) application state failover, and (4) cluster management.

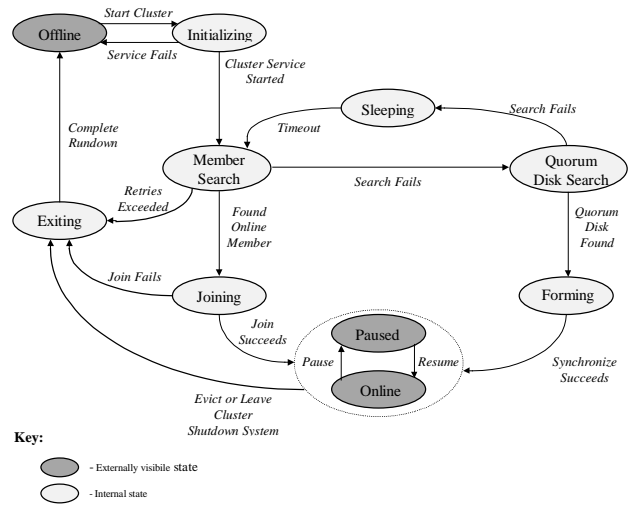

**Figure 1:** State transition diagram for cluster membership.

### 4.1 Cluster Membership Operation

When a cluster node restarts, it can take one of two distinct paths: (1) If there are already active nodes in the cluster, the new node will synchronize with these nodes and join the cluster (i.e. become active). (2) If the node cannot discover any other active cluster nodes, it will try to form a cluster by itself. It will assume it is the first node to start and that other nodes will join later.

The next sections describe the different phases of the membership operation. Section 5 has more details on the membership protocols.

#### Starting a Node

When the node starts, as part of its reboot process it will bring all its local devices online, except for those device that are shared with other nodes. Shared devices may already be controlled by other nodes, so they are only be brought online after the node has joined or formed a cluster. Then the active node negotiates with the other nodes in cluster for device ownership.

The operating system starts the Cluster Service process at node startup. The Cluster Service first enters a discovery phase. The node uses information from its local copy of the cluster configuration database to find the names of the defined nodes (potential cluster members). The node's Cluster Service tries (in parallel) to contact any other Cluster Service at a defined node. If it succeeds in finding an active node, the new node will join the existing cluster. If all the connection attempts time out, the node will try to form a cluster.

### *Joining a Cluster*

If the starting node is able to find an active cluster node, the applicant engages in a startup negotiate with the active node (*sponsor*). First the sponsor validates the authentication credentials of the joining node and checks whether the applicant has a right to join the cluster. If the applicant is a *defined* member of the cluster the sponsor moves to the second phase.

Next the sponsor sends version information of the configuration database and possibly sends database log information to the applicant if changes were made while the applicant was offline. The sponsor then atomically broadcasts information about the applicant to all active nodes members. The active nodes update their local membership information.

Once the applicant is a full member of the cluster and is guaranteed to have access to the correct configuration information, the applicant brings any resources online that it is responsible for and that are not online elsewhere in the cluster.

### *Forming a Cluster*

A node attempts to form its own cluster if it cannot find an active node during the discovery phase. The node uses the local cluster database (registry) to find the address of the quorum resource. The quorum resource holds the master copy of the configuration database and the change logs. The node attempts to attach to the quorum resource. The quorum resource supports an arbitration protocol that assures that at most one node can own the resource. If the node is able to acquire ownership of the quorum resource, the node synchronizes the local cluster database instance with the master copy. When the data in the local database is updated, the node has formed a new instance of the cluster and has become an (the) active member. It can now start bringing shared resources online. Other defined members can now join the newly formed cluster.

### *Leaving a Cluster*

When leaving a cluster, a cluster member sends a *ClusterExit* message to all other members in the cluster, notifying them of its intent to leave the cluster. The exiting cluster member does not wait for any responses but instead immediately proceeds to shutdown all resources and close all connections managed by the cluster software.

The active members gossip about the departed member and update their cluster databases.

### *Node Failure*

To track the availability of the active members in the cluster, all members send periodic heartbeat messages to others and all monitor the network for heartbeat messages (see section 6.1). The communication manager signals a failure suspicion to the Cluster Service when two successive heartbeats have not been received from a particular node. In this case, the Cluster Service starts the *regroup* membership algorithm to determine the current membership in the cluster (see section 5.1). After the new membership has been established, resources that were online at any failed member are brought online at the active nodes, based on the cluster configuration.

### *Node States*

Defined but inactive nodes are *offline*. Active members may be in one of two states: *online* or *paused*. Active members, even paused ones, honor cluster database updates, contribute votes to the quorum algorithm and maintain heartbeats. However, when the node is in paused state it cannot take ownership of any resource groups.

## 4.2 Resource management

The Cluster Service manages resources by invoking a pre-defined set of calls to a *resource control program library* (a dynamically linked library (DLL)) that is provided when the resource type is defined. Central to this is means by which the Cluster Service can monitor the state of the resource. As resources are very diverse it is impossible for the cluster service to have a generic way to manage the state transitions of all resource types. The resource control libraries, which implement the control of the specific resource types (see section 3.3), present a polymorphic state transition mechanism. The resource control libraries for each type hide the complexity of managing state changes for that resource type. This polymorphic design allows for a single Cluster Service to manage varied resource types. One just plugs in a new resource control library for the new resource type.

A resource has five distinct states:
- **Offline**. The (initial) inactive state of the resource, in which it does not provide any service to its clients. In this state the Cluster Service may request the resource to go online. The Cluster Server can bring a resource into offline state by issuing a request while the resource is in online or failed state.
- **Online-pending**. The resource has accepted a request by the Cluster Service to bring its services online. The resource remains in this state while it initializes the service. If the resource fails to initialize it goes into the failed state, otherwise the resource signals the Cluster Service that the resource is online.
- **Online**. The resource is providing services correctly. When the resource is in this state the Cluster Service

uses the resource control library callbacks to check for problems with the resource.
- **Offline-pending**. The Cluster Service has requested the resource to stop offering service and go offline.
- **Failed**. The resource control library has decided that the resource has failed and cannot continue to provide services. This is signaled to the Cluster Service which can decide to restart the resource by bringing it offline and online again (perhaps at another active node.)

The configuration database has a list of possible hosts for a resource and an ordered list of preferred owners. The Cluster Server brings a resource online when any of the possible hosts is available and when all the dependent resources are online.

### 4.3 Resource Migration

A resource group may migrate to another node for many reasons: (1) failure of the original node, (2) failure of the resource at the original node, (3) the resource group prefers to execute at the other node, and (4) the operator requests the group to move. In the first case Cluster Services *pull* the resource groups to the surviving cluster nodes. In the other cases, the owning Cluster Service *pushes* the resource group to the other node.

#### *Pushing a Group*

If a resource fails, the local Cluster Service repeatedly tries to *online* the resource. Failing that, the Cluster Service will optionally move the containing resource group to another node. First all resources in the resource group are taken to the offline state. A new active host node is selected, and the resource group is brought online at the new hosting node by its local Cluster Service. This process is called *pushing* a group to another node. The Cluster Administrator tool and load balancing tools, using the cluster call interfaces, can also initiate a group push.

#### *Pulling a Group*

When an active node fails, its resource groups must be *pulled* to the other active nodes. This process is similar to pushing a resource group, but without the shutdown phase on the failed node. The complication here is determining what groups were running on the failed node and which node should take ownership of the various groups. All nodes capable of hosting the groups determine for themselves the new ownership. This selection is based on node capabilities, the group's preferred owner list, and a simple tie breaker rule, in case the nodes cannot decide which node should be the new host. The replicated cluster database gives all nodes full knowledge of the resource groups on the failed node. Hence, the nodes can determine the new hosts without communicating with one another. Each active node pulls (brings online) the resource groups it now owns.

#### *Fail-back*

A resource group that migrated from its *preferred owner* is not automatically migrated back when the preferred owner rejoins the cluster. Migration back is constrained by the resource group *failback* window described in the Cluster Configuration Database. The failback window indicates how long the new node must be up and running, before the resource group is migrated back to its preferred owner. It also indicates *blackout* periods when failbacks are deferred for cost or availability reasons (migration causes temporary service outage).

### 4.4 Client Access to Resources

NT clients normally access resources using names that look like \\node\service. With MSCS, resources and services migrate among nodes. Clients do not want the service name to change when it migrates -- migration should be transparent to the client. MSCS provides transparency by de-coupling the physical node name from the service name.

Resources are accessed through network names: a NetBIOS name or a DNS name/IP address combination. These names become logical resources that are added to the resource group. Whenever the resource migrates to another node in the cluster, the resource's network name and IP address also migrates as part of its resource group. Consequently, there is an immutable mapping between network names and services. From the client's perspective there are no nodes in the cluster, only services and the network names through which they are accessible: the cluster becomes a single virtual node.

Clients experience a service disruption while a resource group is migrated to a new node. Migration is transparent for connectionless protocols like HTTP and NFS. If the client is connection-oriented, the client must reconnect to the service after the service migrates. This behavior is identical to server failure and reconnect in the monolithic case. Many off-the-shelf client software systems, such as the standard file-system network redirector, already handle this temporary unavailability and reconnect in a transparent manner. Some clients such as ODBC v3 and the SAP R/3 presentation server are cluster aware in that they store sufficient information at the client to reconnect to new instance of the middle tier application servers. These reconnections are not transparent (connection state is lost), but they are automatic.

## 5 Cluster Architecture

The MSCS architecture presents cluster management in three tiers: (1) cluster abstractions, (2) cluster operation, and (3) operating system interaction.

The top tier provides the abstractions described in Section 3: nodes, resources, dependencies, and groups. The important operations features are *resource management*, which controls the local state of resources, and *fail-*

| Table 1: components of the Cluster Service | |
|---|---|
| **Component** | **Functionality** |
| Event processor | Provides intra-component event delivery service |
| Object manager | A simple object management system for the object collections in the Cluster Service |
| Node manager | Controls the quorum *Form* and *Join* process, generates node failure notifications, and manages network and node objects |
| Membership manager | Handles the dynamic cluster membership changes |
| Global Update manager | A distributed atomic update service for the volatile global cluster state variables. |
| Database manager | Implements the Cluster Configuration Database |
| Checkpoint manager | Stores the current state of a resource (in general its registry entries) on persistent storage. |
| Log manager | Provides structured logging to persistent storage and a light-weight transaction mechanism |
| Resource manager | Controls the configuration and state of resources and resource dependency trees. It monitors active resources to see if they are still online |
| Failover manager | Controls the placement of resource groups at cluster nodes. Responds to configuration changes and failure notifications by migrating resource groups |
| Network manger | Provides inter-node communication among cluster members |

*ure management*, which orchestrates responses to failure conditions.

The shared registry allows the cluster service to see a globally consistent view of the cluster's current resource state. The cluster's registry is updated with an atomic update protocol and made persistent using transactional logging techniques. The current cluster membership is recorded in the registry. The membership agreement protocol and failure detection is based on the Tandem multi-computer membership algorithms [4].

The cluster service relies heavily on the Windows NT process and scheduling control, RPC mechanisms, name management, network interface management, security, resource controls, file system, etc. The Cluster Service extends the basic operating system with two new modules. (1) The cluster disk, which implements the Challenge/Defense protocol for shared SCSI disks (section 6.2), and (2) the cluster network module, which implements a simplified interface to intra-cluster communication and implements the heartbeat monitoring.

The core of MSCS consists of 11 components jointly known as the Cluster Service (see table 1). They are combined in a single process, with some communication functionality delegated to the cluster network driver.

The following sections describe some elements of two of the components responsible for the Cluster Service's distributed operation. A detailed description of all components can be found in the technical report [6].

## 5.1 Membership Manager

Membership management is based on the Tandem membership protocol [4]. It maintains consensus among the active nodes: who is active and who is defined. There are two important components to the membership management: (1) the *join* mechanism admits new members into the cluster, and (2) the *regroup* mechanism determines current membership on startup or suspected failure.

### Member Join

Section 4.1 described the operation from the perspective of a joining node. The join algorithm is controlled by he sponsor and has 5 distinct phases for each active node.

The sponsor starts the algorithm by broadcasting the identity of the joining node to all active nodes. It then informs the new node about the current membership and cluster database. This starts the new member's heartbeats. The sponsor waits for the first heartbeat from the new member, and then signals the other nodes to consider the new node a full member. The algorithm finishes with an acknowledgement to the new member

All the broadcasts are repeated RPC's to each active node. If there is a failure during the join operation (detected by an RPC failure), the join is aborted and the new member is removed from the membership.

### Member Regroup

If there is suspicion that an active node has failed, the membership manager runs the *regroup protocol* to detect membership changes. This suspicion can be caused by problems at the communication level, missing heartbeat messages, or power failures.

The regroup algorithm moves each node through six stages. Each node sends periodic status messages to all other nodes, indicating which stage it has finished. None of the nodes can move to the next stage until all nodes have finished the current stage.

1. **Activate**. Each node waits for a local clock tick so that it knows that its timeout system is working. After that the node starts sending and collecting status messages. It advances to the next stage if all active nodes have responded, or when the maximum waiting time has elapsed.
2. **Closing**. This stage determines whether partitions exist and whether the current node is in a partition that

should survive. A partition survives if *any one* of the following conditions is satisfied
1. The new membership contains more than half the original membership.
2. (a) the new membership has exactly half the original members, and (b) there are at least two members in the current membership, and (c) this membership contains the tie breaker node that was selected when the cluster was formed.
3. (a) the original membership contained exactly two members and (b) the new membership only has one member, and (c) this node owned the quorum disk in the previous group.

If the new group survives, the new members select a tiebreaker node to use in the next regroup. This tiebreaker then checks the connectivity information received from all nodes to ensure that the surviving group is fully connected. (If not it prunes the partially connected members nodes). It then announces the new membership to all members. All now move to stage 3.

3. **Pruning**. All nodes that have been pruned for lack of connectivity halt in this phase. All others move forward to the first cleanup phase.
4. **Cleanup Phase One**. All surviving nodes install the new membership mark the nodes that did not survive the membership change as inactive, and inform the cluster network manager to filter out messages from these nodes. Each node's Event Manger then invokes local callback handlers to announce the node failures.
5. **Cleanup Phase Two**. Once all members have indicated that the Cleanup Phase One has been successfully executed, a second cleanup callback is invoked to allow a coordinated two-phase cleanup. Once all members have signaled the completion of this last cleanup phase they move to the final state.
6. **Stabilized**. The regroup has finished.

There are several points during the operation, where timeouts can occur. These timeouts cause the regroup operation to restart at phase 1.

### *5.2 Global Update Manager*

Many components of the NT Cluster Service need to share volatile global state among nodes. The algorithm used by the Global Update Manager is a variant of the Tandem GLUP protocol [2]. It is an atomic multicast protocol guaranteeing that if one surviving member in the cluster receives an update, all surviving members eventually receive the update, even if the original sender fails. It also guarantees that updates are applied in a serial order.

### *Locker Node*

One cluster node, dubbed the *locker node,* is assigned a central role in the Global Update Protocol. Any node that wants to start a global update first contacts the locker. The locker node promises that if the sender fails during the update, the *locker* (or it's successor) will take over the role of sender and update the remaining nodes. Once the sender is finished updating all the members in the cluster it sends the *locker* node an *unlock* request to indicate the protocol terminated successfully.

### *Node Updates*

Once a sender knows that the locker has accepted the update, the sender RPCs to each active node (including itself) to install the update. The nodes are updated one-at-a-time in a node-ID order starting with the node immediately following the locker node, and wrapping around the ID's up to the node with ID preceding the locker's. Once the update has been installed at all nodes, the locker is notified of the completion.

### *Failures*

The protocol assumes that if all nodes that received the update fail, it is as if the update never occurred. The remaining nodes do not need to recover such updates. Examples of such failures are (1) sender fails before locker accepts update, or (2) sender installs the update at the locker, but both sender and locker node fail after that.

If the sender fails during the update process, the locker reconstructs the update and sends it to each active node. Nodes that already received the update detect this through a duplicate sequence number and ignore the duplicate update.

If the sender and locker nodes both fail after the sender managed to install the update at any node beyond the locker node, a new locker node will be assigned. This new locker node will always be the next node in the update list. Given the way the updates are ordered, this node must have already received the update. If the sender managed to install the update past the locker node, it did starting at the node immediately following the locker node. The new locker will complete any update that was in progress using the saved update information. To make this work, the locker allows at most one update at a time. This gives a total ordering property to the protocol -- updates are applied in a serial order.

## 6   Support Components.

The Cluster Service uses several support components unique to MSCS: the cluster network component, the cluster disk driver, the event logger, and the time service.

### *6.1   Cluster Network*

The Cluster Network component provides the cluster service with:
1. A uniform interface to communicate with other nodes, independent of the network infrastructure.
2. Predictable processing of *I'm-Alive* heartbeat messages
3. Node failures detection based on heartbeats.
4. Network and interface failure detection.

### *Heartbeat Management*

The cluster network keeps an active-node connectivity vector. For each active node, the network manager keeps a list of interfaces that can reach that node. Each node periodically sends a heartbeat message to each other active node over each of these interfaces. When a heartbeat arrives from an active node, the sender's local timeout counter for that node is reset, and its heartbeat sequence number is recorded. Duplicates arriving over alternate interfaces are ignored for the node's alive count, but do test the network for latent failures. If, after a certain period (currently 2 heartbeat periods), no messages have arrived over a certain interface or from a particular host, failure suspicions are generated.

### 6.2 Cluster Disk driver

MSCS provides support for the disks connected to a shared SCSI bus. Multiple nodes in the cluster can be connected to the same SCSI bus. At any time, each disk is "owned" by one of the nodes, but each node can own some disks on the bus. This allows dynamic ownership and failover of disks among cluster members.

The Cluster Service uses the shared disk mechanism to implement the Cluster Quorum Resource. The quorum resource has two roles: (1) it breaks ties when exactly half of the nodes are trying to form a cluster, and (2) it stores the cluster database and log information across cluster failure periods. By using the SCSI challenge-defense protocol, the cluster service arbitrates for the ownership of the quorum resource.

### 6.3 Cluster Wide Event Logging

Event logs are an important tool for NT server administrators. Logs track the execution state and potential problems of NT devices, services, and the applications running on the node. The administrator uses an event viewer to display the logs. In a cluster there is no longer a clear association between node and the applications and services running on the node, as such it is complex for to track nodes and services using the traditional mechanisms.

The Cluster Service extends the event log mechanism by enabling administrators to view a single event log containing all the events in the cluster, even if the node that reported the event is currently down.

To implement this global event log, the local event log mechanism forwards local events to the local Cluster Services. Events reported to the Cluster Service are sent via RPC to all other nodes in the cluster, where they are appended to the local event log files.

### 6.4 Time Service

The Cluster Service ensures that the clocks at the nodes in the cluster never drift apart more than the shortest time it takes to failover a resource. This ensures that resources that failover between nodes see a monotonically increasing local clock. The Cluster Service uses the standard NT Time System Service, but uses its resource control library to dynamically update the registry information to match the primary time source within the cluster. This allows all clocks to be synchronized with universal time.

## 7 Virtual Servers

*Virtual NT servers* extend the resource group concept to provide a simple abstraction for applications and administrators. Applications run within a *virtual server* environment. This environment provides the application with the illusion that it is running on a virtual NT node with virtual services, a virtual registry, and with a virtual name space. When an application migrates to another node, it appears to the application that it restarted at the same virtual NT node.

The Virtual Server environment provides applications, administrators, and clients with the illusion of a single, stable environment -- even if the resource group migrates.

One benefit of virtual servers is that many instances of an application can be executed on a single node, each within its own virtual server environment. This allows two SQL Servers or two SAP environments to execute as two virtual servers on one physical NT node.

### 7.1 Client Access

A virtual server resource group requires a *node name* resource (NetBios and DNS), and an *IP address* resource. Together, these present consistent naming for the clients. The virtual server name and IP address migrate among several physical nodes. The client connects using the virtual server name, without regard to the physical location of the server.

### 7.2 Server Encapsulation

Each virtual server environment provides a name space and configuration space separated from other virtual servers running at the same node: registry access, service control, named communication (pipes), and RPC endpoints. This allows two instances of an application service running on the same node but in separate virtual service environments not to clash in the access of configuration data or internal communication patterns.

To provide this transparency, three features were added to NT:

- **Virtual server naming**. System services (such as GetComputerName) return the network name associated with the virtual server instead of the host node name.
- **Named pipe re-mapping**. When an application service consist of several components that use interprocess communication to access each other's services, the communication endpoints must be named relative to the virtual server. To achieve this a name remapping facility for the named pipe interface was implemented. Named pipe names are translated from \\virtual_node\service to \\host\$virtual_node\service.
- **Registry replication**. The Windows NT registry stores most application configuration data. To allow applications that run in separate virtual servers to run on the same host node, registry trees must be remapped to virtual-server local trees. nt state in the registry, parts Each unique tree represents a single virtual server and is internally dependent on the name associated with the virtual server. When the virtual server migrates, the local tree is rebuilt from the logs on the quorum device.

Although virtual servers are a clean abstraction for encapsulating applications, the implementation is extremely difficult. It is difficult to know all the dependencies on node specific resources, as applications often make use of dynamic loadable libraries in ways that introduce new naming dependencies.

# 8 Experience

Most major NT Server applications have been modified to use MSCS. The next sections describe three of them: Microsoft SQL Server, Oracle Parallel and Failsafe servers, and the SAP R/3 business system.

## 8.1 Microsoft SQL-Server

SQL Server 6.5 is a client/server relational database system consisting of a main server process and a helper process called the executive. The server process manages a collection of databases that in turn map down to the NT file system. The server is a free-threaded process that listens for SQL requests and executes them against the database. It is common to encapsulate the database with stored procedures that act on the database. Client requests invoke these procedures, written in the TransactSQL programming language. The procedures execute a program with control flow and calls to read and write the database. These procedures can also access other NT services (e.g. Mail is often used for operator notification.)

The SQL executive performs housekeeping tasks like launching periodic jobs to backup or replicate the database. The replication is quite flexible. It is often used to create a replica of the database that can be used in case the primary system fails.

### SQL Server Database Failover

In 1996, Microsoft shipped a version of SQL Server that provided database failover. This design predated MSCS. In that design, two SQL Servers were configured on two NT nodes with disks shared between them. Each database, (a self-describing collection of SQL tables, with its own transaction log file) was configured as a subset of the shared disks. SQL Server was able to migrate a database from one server to another. The ODBC client protocol was aware of the different server names and was enhanced to select the second server for a database if the first one was not available.

This design "worked" but it had some unexpected problems. Fundamentally, it was very hard to configure correctly, especially at the server side. A database that moves between two SQL servers finds itself with two sets of registry entries that determine how the server (and the database) should behave. Administrators quickly learned that they needed to keep the two servers *identical*. One important difficulty is that the administrator may not realize that some operation is modifying the registry. Similarly, security on the two servers had to be kept identical. Stored procedures often reside in a server-global database. Administrators discovered that they needed to exactly replicate these procedures at each server so that when the database failed over to the second server, clients would be able to find the same procedures. SQL Server's replication software and many utilities did not use the ODBC interface and so did not failover. This effectively disabled replication -- a major loss of functionality.

### SQL Server Failover under MCSC.

Using MSCS it was possible to move away from database failover and redesign SQL Server to use server failover as the mechanism for high-availability. A SQL Server resource group is configured as a virtual NT server with a virtual registry, virtual devices, virtual services, virtual name, virtual IP address, and whatever else is needed to create the fiction that it is a virtual NT node. Clients connect to the server running on this virtual node. The server application thinks it is running on this virtual server and the virtual server can migrate among physical servers.

This design allows a two-node MSCS cluster to have two or more high-availability SQL Servers. The clients always see the same server name, even as the server migrates from node to node. The server sees the same environment, even as it migrates. Administrators, who are really just clients, administer virtual servers, just like real servers (no new concepts have been added). Replication and other mechanisms just work. Again, it is as if nothing has changed -- the server is just virtualized. This simple design has proved to be very easy to understand and to explain to users.

## 8.2 Oracle Database Servers

Oracle Corporation's Oracle Parallel Server (OPS) runs on most cluster environments. OPS relies on the *shared disk* model for its cluster technology, using a distributed lock manager (DLM) to coordinate disk accesses. Each instance of the OPS database server can access all databases stored on the shared disk array. Whenever an instance of the server fails, a surviving OPS runs a recovery operation to recover the database.

*Classic* OPS running on an MSCS cluster uses the same model. Consequently, the shared disks are not resources managed by MSCS. OPS just uses MSCS to track cluster organization and membership notifications.

Oracle developed the Oracle Fail-Safe product to integrate more closely with MSCS. Fail-Safe uses the MSCS *shared nothing* model and *data partitioning*. Each Fail-Safe server instance manages one or more databases.

A Fail-Safe server instance runs as a service (process) at each node in the cluster. It performs database monitoring (interacting with the MSCS resource monitors) for each database at that node, and it maintains the dynamic configuration of the client/server interface modules.

The Oracle Fail-Safe server is a *container service* in the sense that a single server process at a node supports many resource groups. Resource groups can migrate to and from the container service. Microsoft's SQL Server by contrast is a *process-service*: there is one resource group per process, and new processes are created when the resource migrates.

Each Oracle Fail-Safe database is stored on one or more shared disks, but conforming to the MSCS model, the database is only accessible to the Oracle server that controls the disk as a cluster resource. Each instance of a Fail-Safe database is an MSCS virtual server. Upon failure, the virtual server (disk, database, IP address, and network name) migrates to another active node. Once the resources are brought on-line, the local Fail-Safe server is notified of the new resources. The Fail-Safe server initiates recovery and begins offering database access.

Clients access a database through the network names and IP addresses associated with the resource group that holds each database. When a database fails, the client reconnects to the database server under the same name and address, which have moved to the surviving node. Application builders are encouraged to maintain transaction state information, so that after a failure any in-flight transactions can be resubmitted.

### 8.3    SAP R/3

SAP AG offers a scalable system for business applications. It is the world's largest supplier of business solutions. SAP builds its systems around the sophisticated middleware system SAP R/3. R/3 is a three-tier Client/Server system, separating presentation, application and data storage into clearly isolated layers. Although all the business logic is in the application tier of the system, there is no persistent state at the application servers. This separation allows application servers to be added to system to achieve higher processing capacity and availability. Load balancing and availability management of the parallel application servers is through dedicated R/3 management tools.

All persistent state of R/3 is maintained in the database tier where a $3^{rd}$ party database is used for data storage. To facilitate the use of different database vendors, the system avoids the storage of non-portable elements such as database stored procedures. There are 3 components of which only a single instance can be active in the system, and which are thus a single point of failure: the database, the message server and the enqueue server. The failure of any of these servers will bring the complete system to a halt.

These services are made highly available with MSCS.

Given that only single instances of these servers can be active in the system, partitioned failover is used to organize the cluster. During normal operation, one node of the cluster hosts the database virtual server and the other provides the SAP middleware and enqueue virtual server that are combines the application components into a server dubbed the Central Instance (CI). Upon failure of either node, the failed virtual server migrates to the surviving node.  The application components are all placed in a joint resource group, with dedicated resource libraries managing each of the components. The resources are organized in several complex dependency trees. SAP application migration is relatively simple as no persistent state is kept at this tier.

The application servers in SAP R/3 are "failover aware", in that they can be temporarily disconnected from the database or message/enqueue server. After a waiting period they try to reconnect. A failure in the database tier is transparent to the user, as the application servers mask the potential transaction failure that was the result of the failure. Migration of the application server is handled at the client by establishing a new login-session at the new node hosting the application server.

## 9    Future Directions

We commented that MSCS is still in a first phase. The second phase will focus on scalability and extending availability management. The scaling effort has two major components: First there is cluster software itself, where scalability of the algorithms used in the distributed operation is critical to growing clusters to larger numbers of nodes [7]. Secondly, we are exploring the introduction of cluster aware mechanisms for use by developers of sophisticated servers that might exploit cluster-style parallelism. In support of such an approach it would be necessary to provide programming interfaces for cluster membership and communication services, similar to services in Isis [1]. Server developers would then use these to build scalable applications.